\documentclass[aps,prl,twocolumn,groupedaddress,floatfix,notitlepage,nofootinbib]{revtex4-2}

\usepackage{graphicx}
\usepackage{bm}				
\usepackage{amsfonts}
\usepackage{amssymb}
\usepackage{amsmath}
\usepackage{ulem} 
\usepackage{graphicx}		
\usepackage{xcolor, soul}	
\usepackage{verbatim}
\usepackage[T1]{fontenc}
\usepackage{pifont}
\usepackage{multirow}
\usepackage{notes2bib}
\usepackage{mathrsfs}
\usepackage{natbib}
\usepackage{appendix}

\usepackage[colorlinks=true,linkcolor=black,citecolor=blue,urlcolor=blue,hypertexnames=false]{hyperref}

\definecolor{darkgreen}{rgb}{0,0.5,0}

\newcommand{\ourtitle}{
Geometric Spin Degeneracy in Spin-Orbit-Free Compensated Magnets
}

\begin{document}

\title{\ourtitle}	
	
    \author{Seung Hun Lee$^{1,2,3,4,*}$}
    \author{Yuting Qian$^{1,2,3,*,\dagger}$}
    \author{Xi Dai$^{4}$}
    \author{Bohm-Jung Yang$^{1,2,3,\ddagger}$}
    \affiliation{$^1$Department of Physics and Astronomy, Seoul National University, Seoul 08826, Korea}
    \affiliation{$^2$Center for Theoretical Physics (CTP), Seoul National University, Seoul 08826, Korea}
    \affiliation{$^3$Institute of Applied Physics, Seoul National University, Seoul 08826, Korea}
    \affiliation{$^4$Department of Physics and Center for Complex Quantum Systems, The University of Texas at Austin, Austin, Texas, 78712, USA.}
    \affiliation{$^5$Department of Physics, Hong Kong University of Science and Technology, Clear Water Bay, Kowloon 999077, Hong Kong}
	
    \date{\today}
	
    \begin{abstract}
    Compensated magnets with vanishing net magnetization can exhibit both pronounced spin splitting and unconventional band degeneracies. In altermagnets, such degeneracies are enforced by crystal and magnetic symmetries. In compensated ferrimagnets, however, they may arise even in the absence of the corresponding symmetry protection, raising a fundamental question about the origin of spin degeneracy in spin-orbit-free magnetic systems. Here, we develop a theoretical framework for spin-orbit-free compensated magnets in which spin degeneracies are protected by geometric constraints rather than by spin symmetry. We show that zero net magnetization imposes a strong condition for the emergence of nodes formed by formally spin-degenerate bands, even when no conventional spin symmetry is present. Our analysis, applicable in the weak-interaction regime, identifies a general mechanism for spin degeneracy beyond group-theoretical protection. The framework accounts for the unconventional spin degeneracies recently reported in compensated ferrimagnets and provides a unified description of band degeneracies across a broad class of magnetic phases with negligible spin-orbit coupling.
    \end{abstract}

    \maketitle
    \begingroup
    \renewcommand{\thefootnote}{\fnsymbol{footnote}} 
    \footnotetext[1]{These authors contributed equally to this work.}
    \footnotetext[2]{yuting\_qian@snu.ac.kr}
    \footnotetext[3]{bjyang@snu.ac.kr}
    \endgroup

\begin{figure}[h!]
	\centering
	\includegraphics[width=1\linewidth]{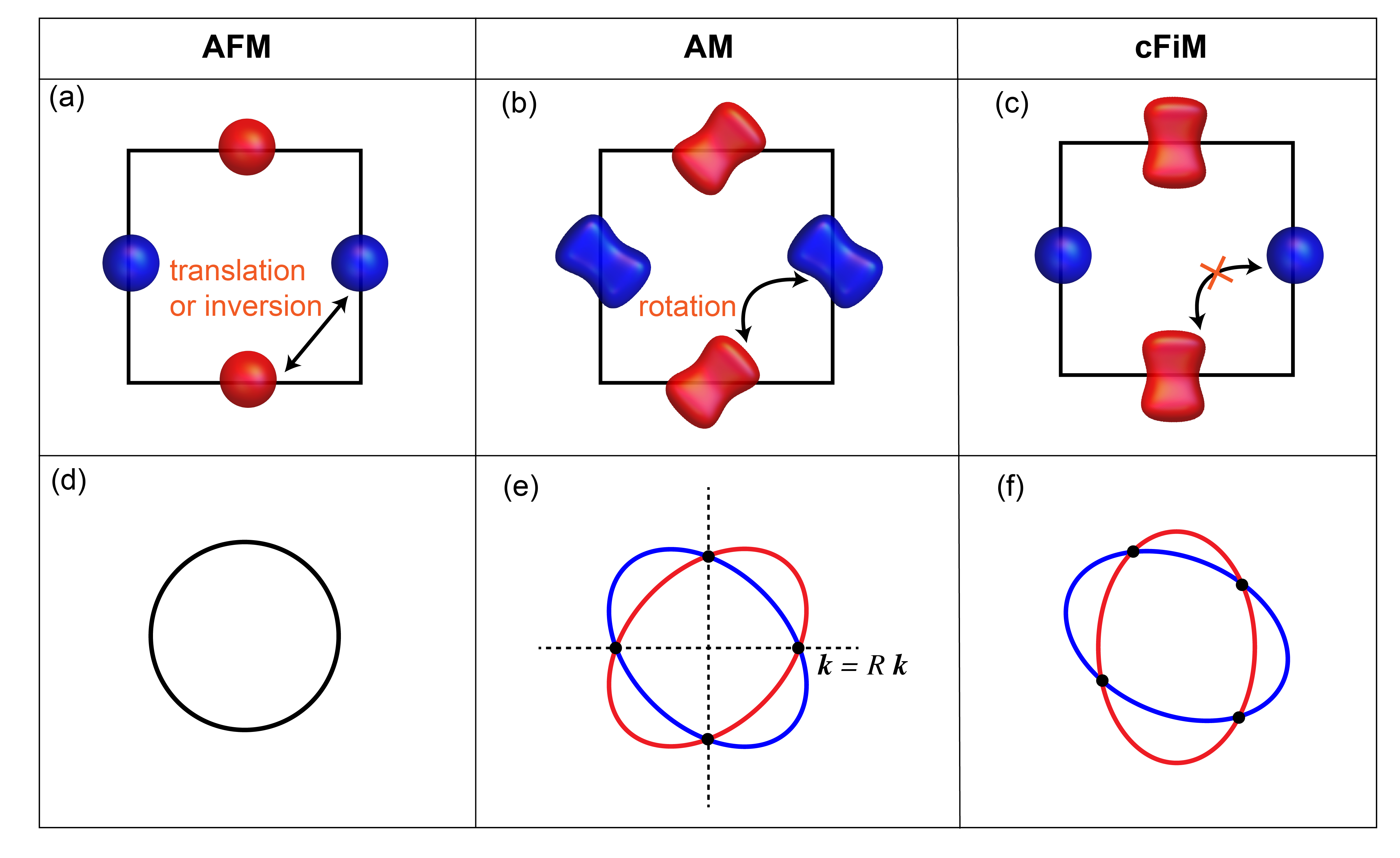}
	\caption{\textbf{Comparison of nodal structures in different types of spin-orbit-free compensated magnets.}
    The red and blue surfaces represent magnetic atoms with opposite spins.
    (a) In conventional antiferromagnets (AFMs), sublattices with opposite spins are related by translation or inversion symmetry.
    (b) In altermagnets (AMs), sublattices with opposite spins are not related by translation or inversion, but by other crystalline symmetries, such as rotation.
    (c) In compensated ferrimagnets (cFiMs), sublattices with opposite spins are not related by any symmetry.
    (d–f) Nodal structures of AFM, AM, and cFiM, respectively. The red and blue lines represent Fermi surfaces of spin-up and spin-down electrons, respectively. Black lines and dots indicate degenerate points between spin-up and spin-down Fermi surfaces.
    }\label{schematics}
    \end{figure}

    \textit{Introduction.}---
    Broken time-reversal symmetry allows spin-up and spin-down electrons in magnetic metals to behave differently, thereby shaping their electronic properties. In ferromagnets (FMs), spin splitting gives rise to spin-dependent transport phenomena and unconventional superconducting phases~\cite{jungwirth2016anti,baltz2018}. In conventional antiferromagnets (AFMs), by contrast, such effects are typically absent because the bands remain fully spin degenerate [Fig.~\ref{schematics}(a,d)]. Recently, however, new classes of magnetic materials have emerged that combine compensated magnetic moments with spin-split band structures. A prominent example is altermagnets (AMs), collinear antiferromagnets with broken Kramers degeneracy~\cite{prxeditor,vsmejkal2022emerging}, which have opened new avenues for spin-dependent phenomena in materials with zero net magnetization.

    Compensated ferrimagnets (cFiMs)~\cite{de1991half,leuken1995,wurmehl06,ogura2006half,cmferriprl,cmferriprb,geprags2016origin,half17,ferri2022,liuchengcheng} provide another such platform with potentially useful spin-dependent functionality arising from large spin splitting. Their magnetic sublattices carry unequal local moments, yet the net magnetization vanishes either under specific conditions, such as at a compensation temperature~\cite{geprags2016origin}, or identically because of electronic constraints, for example those associated with the Luttinger theorem~\cite{prxeditor,liuchengcheng}.
    These systems are also attractive for applications~\cite{Warren96,spintronics, rosenberg2018,finley2016spin,caretta2018,cai2020,MnF2,Ferri24} because they produce no stray field and are less sensitive to external magnetic fields. Unlike AFMs [Fig.~\ref{schematics}(a,d)] and AMs [Fig.~\ref{schematics}(b,e)], where symmetries constrain spin degeneracy, cFiMs generally lack a well-defined symmetry relating the spin-up and spin-down sectors, rendering conventional group-theoretical approaches inapplicable~\cite{jiang2024enumeration,xiao2024spin,chen2024enumeration}.
    Nevertheless, cFiMs can exhibit even larger spin splitting than altermagnets while retaining spin degeneracies in parts of momentum space [Fig.~\ref{schematics}(c,f)]. Whether these degeneracies are accidental or reflect a deeper underlying principle remains unclear, calling for a framework beyond symmetry-based analysis.

    In this work, motivated by the spin degeneracies observed in cFiMs, we develop a geometric theory of spin degeneracy in spin-orbit-free magnets that does not rely on spin symmetry. We show that vanishing net magnetization imposes a strong condition for the existence of nodes formed by formally spin-degenerate bands, even in the absence of conventional spin symmetry. Our theory applies in the weak-interaction regime, where the electron-electron interaction is small compared with the gap separating the bands of interest from adjacent bands in the nonmagnetic limit. Importantly, the framework is not restricted to ferrimagnetic systems, but extends more broadly to magnetic materials with negligible spin-orbit coupling, providing a unified understanding of band degeneracies across a wide range of magnetic phases.
     
    \textit{Geometric condition for degeneracy in spin-orbit-free magnets.}\label{spliting}---
    To describe the spin splitting in spin-orbit-free magnets, we introduce \textit{effective Zeeman field}, whose precise definition is given below.
    We define a spin-degenerate band in a given non-magnetic system as a parent band, and refer to the two spin-split bands that emerge from this parent band upon magnetic ordering as a \textit{formal} Kramers pair.
    Below, we show that vanishing net magnetization is an important condition that guarantees the existence of nodes in the band structure composed of the formal Kramers pair.
    
    Let us consider a tight-binding Hamiltonian $\hat H$ that describes a spin-orbit-free magnet. In momentum space, we write $\hat H=\sum_{\bm k} \hat{\bm c}_{\bm k}^\dagger H(\bm k)\hat{\bm c}_{\bm k}$ where we take the basis $\hat{\bm c}_{\bm k}^\dagger=(\hat c_{\bm k1\uparrow}^\dagger,\hat c_{\bm k2\uparrow}^\dagger,...,\hat c_{\bm kN\uparrow}^\dagger,\hat c_{\bm k1\downarrow}^\dagger,\hat c_{\bm k2\downarrow}^\dagger,...,\hat c_{\bm kN\downarrow}^\dagger)$ and $H(\bm k)=H_0(\bm k)\otimes\sigma_0+H_m-\mu I_{2N}$.
    $H_0(\bm k)$ is the kinetic part of $H(\bm k)$ and describes the spin-independent electron hopping with an identity matrix $\sigma_0$ in the spin space.
    $\mu$ is the chemical potential and $I_{2N}$ is an $2N\times 2N$ identity matrix.
    $H_m=\sum_{\alpha=x,y,z} h_m^\alpha\otimes\sigma_\alpha$ describes the coupling of electrons to local spin moments at $N$ sublattice sites in a unit cell.
    $h_m^\alpha$ is a diagonal matrix in the form of $h_m^\alpha=\textrm{diag}(a_1^\alpha,a_2^\alpha,...,a_N^\alpha)$ in which the real diagonal entries indicate the local spin's $\alpha$-component at each sublattice site.

    $H_0(\bm k)$ can be diagonalized by a unitary transformation as $U_0^\dagger(\bm k) H_0(\bm k) U_0(\bm k)=\textrm{diag}(\varepsilon_1(\bm k),\varepsilon_2(\bm k),...,\varepsilon_N(\bm k))$, where $U_0(\bm k)=(|u_1(\bm k)\rangle,|u_2(\bm k)\rangle,...,|u_N(\bm k)\rangle)$ and $H_0(\bm k)|u_i(\bm k)\rangle=\varepsilon_i(\bm k)|u_i(\bm k)\rangle$.
    Let us assume that the bandwidth of each band $\varepsilon_i(\bm k)$ is much larger than $a_n^\alpha$. 
    To analyze the contribution of $H_m$, we project it onto the band basis of $H_0(\bm k)$ and obtain an effective Hamiltonian
    \begin{align}
        H_{i,\textrm{eff}}(\bm k)&\equiv
        \begin{pmatrix}
            \langle u_{i,\uparrow}|H|u_{i,\uparrow}\rangle & \langle u_{i,\uparrow}|H|u_{i,\downarrow}\rangle\\
            \langle u_{i,\downarrow}|H|u_{i,\uparrow}\rangle & \langle u_{i,\downarrow}|H|u_{i,\downarrow}\rangle
        \end{pmatrix}\nonumber\\
        &=\varepsilon_i(\bm k)\sigma_0+\bm f_i(\bm k)\cdot\bm \sigma
        \label{eqn:Heff}
    \end{align}
    for each $i$-th band, where $\bm{\sigma}$ is the vector of spin Pauli matrices with components $\sigma_\alpha ({\alpha=x,y,z})$.
    Since $H_m$ transforms as $U^\dagger(\bm k) H_m U(\bm k)=\sum_\alpha U_0^\dagger (\bm k) h_m^\alpha U_0(\bm k)\otimes\sigma_\alpha$ under the unitary matrix $U(\bm k)=U_0(\bm k)\otimes\sigma_0$, 
    the influence of $H_m$ on the $i$-th band can be captured by a momentum-dependent \textit{effective Zeeman field} (EZF) $\sum_\alpha f_i^\alpha(\bm k)\sigma_\alpha$, where $f_i^\alpha(\bm k)\equiv[U_0(\bm k)^\dagger h_m^\alpha U_0(\bm k)]_{ii}=\sum_{n=1}^N a_n^\alpha u_{in}^*u_{in}=\sum_{n=1}^N a_n^\alpha\|u_{in}\|^2$.
    Here, $f_i^\alpha(\bm k)$ quantifies the effective Zeeman splitting experienced by an electron in the $i$-th band, determined by the electron wavefunctions in the non-magnetic limit.
    Specifically, the quantity $\|u_{in}\|^2$, which represents the sublattice-resolved probability weight in the wavefunction of the $i$-th band, directly determines how much of the local spin moment at sublattice $n$ contributes to the spin expectation value of the wavefunction.

   \begin{figure}[!t]
	\centering
	\includegraphics[width=1\linewidth]{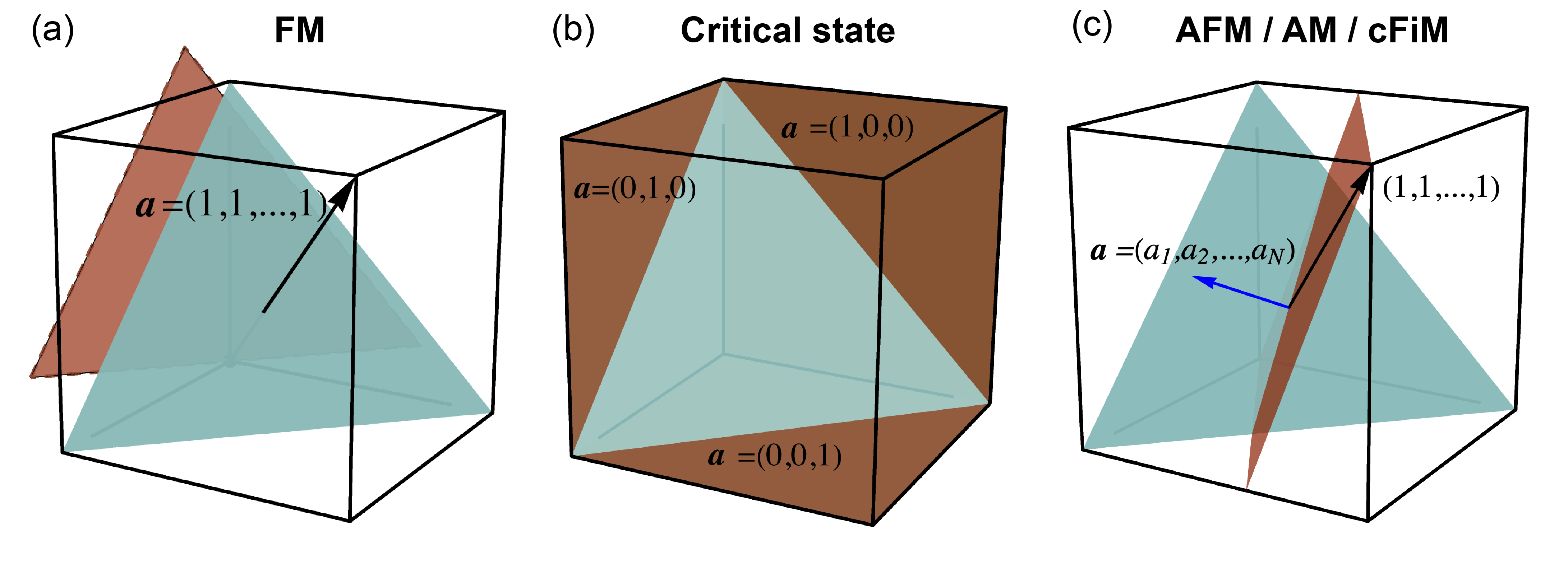}
	\caption{\textbf{Geometric approach to band degeneracies in representative magnetic states.}
    Hilbert polygon (green plane) and different ZEZF planes (brown plane) with normal vector $\bm a$. (a) FM. Green and brown planes are parallel. (b) An intermediate critical state with $\bm a=(1,0,0)$, or $(0,1,0)$, or $(0,0,1)$. (c) AFM/AM/cFiM with zero magnetization. Here, FM: ferromagnet; AFM: antiferromagnet; AM: altermagnet; cFiM: compensated ferrimagnet.
    }\label{nearferro}
    \end{figure}
    
    With the EZF defined, we can now determine the spin-degenerate momenta. The spin degeneracy appears within the formal Kramers pair when the EZF vanishes even after the introduction of magnetism.
    We can find the momenta at which EZF vanishes through a geometric approach as follows.
    We first define a map from the Hilbert space $\mathcal{H}$ to the real coordinate space $\mathbb{R}^N$, where $N$ is the dimension of the sublattice degree-of-freedom, as $\mathcal{H} \rightarrow \mathbb{R}^N:|u_{i}(\bm k)\rangle\mapsto V_{i}$ where $V_{i}=(V_{i1},V_{i2},...,V_{iN})=(\|u_{i1}\|^2,\|u_{i2}\|^2,...,\|u_{iN}\|^2)$.
    From the EZF equation introduced above, we find that the condition for vanishing EZF is given by $f_i^\alpha=\sum_n a_n^\alpha V_{in} =0$ for $\alpha=x,y,z$. In $\mathbb{R}^N$, these are equations of $(N-1)$-dimensional hyperplanes containing the origin. We name these hyperplanes as zero EZF (ZEZF) planes, whose normal vectors are $\bm a^\alpha=(a_1^\alpha,a_2^\alpha,...,a_N^\alpha)$.
    For each spin direction $\alpha=x,y,z$, we obtain one corresponding hyperplane in $\mathbb{R}^N$.
    
    In addition to the set of ZEZF hyperplane equations that correspond to different $\alpha$s, we have a constraint related to the normalization of $|u_i(\bm k)\rangle$.
    That is, $\sum_{n}\|u_{in}\|^2=\sum_{n} V_{in}=1$.
    In the space of $V_i$, this constraint is represented by a regular $N$-polygon whose vertices are at $(1,0,...,0)$, $(0,1,...,0)$,..., $(0,...,0,1)$, respectively. We name this polygon as the Hilbert polygon, since it is nothing but the image of the full Hilbert space generated by the map.
    The Hilbert polygon and ZEZF planes with corresponding normal vector $\bm a$ are schematically illustrated in Fig.~\ref{nearferro} for different magnetic configurations.
    Finding spin degeneracy amounts to identifying solutions of coupled equations $f_i^\alpha=0$ ($\alpha=x,y,z$) and $\sum_{n}V_{in}=1$, which is equivalent to stating that EZF vanishes at the intersection of these ZEZF planes and the Hilbert polygon.
    Next, by taking the kagome lattice as an example, we demonstrate that the zero net magnetization condition, i.e., $\sum_i a_i^\alpha=0$, serves as a strong condition to guarantee the existence of such solutions.

    \textit{Kagome lattice with collinear magnetism.}\label{kagome_collinear}---
    To demonstrate our theory, we consider a single $s$-orbital kagome lattice model with nearest-neighbor hopping described by the Hamiltonian
    \begin{equation}
        H_0(\bm k)=-t
        \begin{pmatrix}
        0 & \cos\frac{\bm k\cdot\bm e_1}{2} & \cos\frac{\bm k\cdot\bm e_3}{2} \\
        \cos\frac{\bm k\cdot\bm e_1}{2} & 0 & \cos\frac{\bm k\cdot\bm e_2}{2} \\
        \cos\frac{\bm k\cdot\bm e_3}{2} & \cos\frac{\bm k\cdot\bm e_2}{2} & 0 \\
        \end{pmatrix},
    \end{equation}
    where $t$ is the hopping parameter, $\bm k=(k_x,k_y)$ and the lattice vectors $\bm e_i:(1,0)$, $(-1/2,\sqrt{3}/2)$, $(-1/2,-\sqrt{3}/2)$ for $i=1$, 2, 3, respectively.
    
    Let us first consider a ferromagnetic ordering by using $H_m$ in the form of $m I_{3}\otimes\sigma_\alpha$ where $I_3$ is the $3$-dimensional identity matrix corresponding to $N=3$ sublattice sites.
    This implies that the ZEZF plane has the normal vector $(1,1,1)$.
    Consequently, a ferromagnetic ZEZF plane passing through the origin is always parallel to the Hilbert polygon as shown in Fig.~\ref{nearferro} (a).
    As a result, these ZEZF planes never intersect with the Hilbert polygon, and spin-degenerate nodes within a formal Kramers pair are therefore forbidden in ferromagnets.
    As the magnetic ordering pattern continuously evolves from the purely ferromagnetic limit, the ZEZF plane also tilts gradually, and across a critical intermediate state described in Fig.~\ref{nearferro} (b), it will begin to intersect the Hilbert polygon.
    When the system finally reaches the magnetic configuration with vanishing net magnetization, the corresponding ZEZF plane always intersects the Hilbert polygon, as depicted in Fig.~\ref{nearferro} (c).
    
\begin{figure}[t!]
	\centering
	\includegraphics[width=1\linewidth]{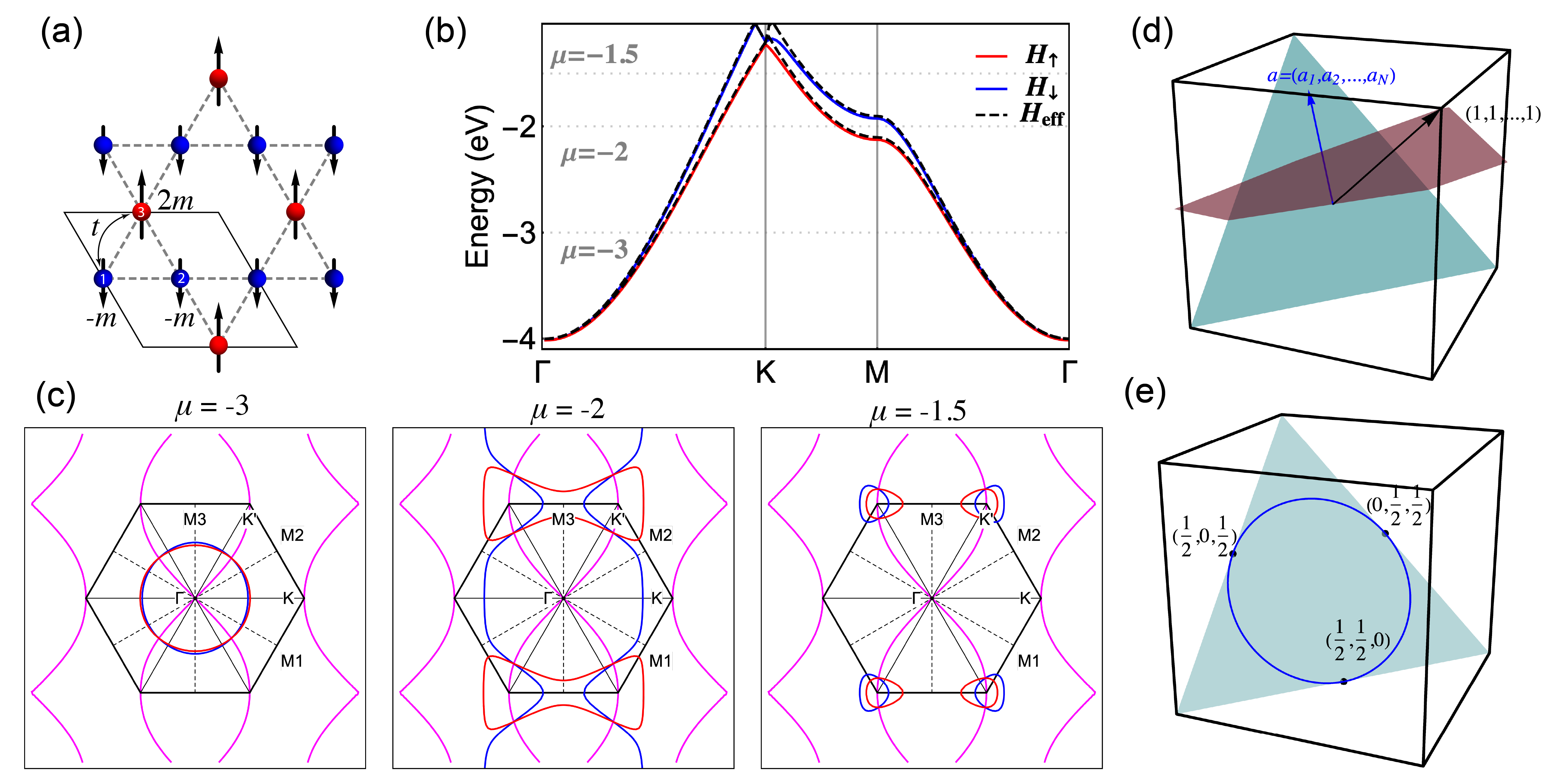}
	\caption{\textbf{A cFiM model on the kagome lattice.}
        (a) The tight-binding model for a collinear spin-orbit-free magnet on a kagome lattice with $h_m^x=h_m^z=\textrm{diag}(0,0,0)$ and $h_m^y=\textrm{diag}(-m,-m,2m)$. $t$ describes the nearest-neighbor hopping strength. 
        (b) Energy dispersion of the two lowest bands obtained from the tight-binding model in (a) with $t=1.0$ and $m=0.4$. Red (Blue) band structures are for spin-up (spin-down) bands of $H(\textbf{k})$, and the black dotted lines indicate the band structure of the effective Hamiltonian $H_\textrm{eff}(\textbf{k})$. 
        (c) The spin-up (red) and spin-down (blue) Fermi surfaces for different chemical potentials with the magenta lines representing the ZEZF momenta.
        (d) The ZEZF plane (brown) and the Hilbert polygon (green) for a collinear magnet on the kagome lattice, with the normal vector $a^y=(-1,-1,2)$.
        (e) The blue line represents the incircle, corresponding to the image of all possible linear combinations of the two degenerate eigenstates of $H_0(\mathbf{k})$ at the $K$ point.
    }\label{fig:kagome}
    \end{figure}

    To demonstrate the evolution of the geometric conditions, we next consider compensated magnetic configuration described by $h_m^x=h_m^z=0$ and $h_m^y=\textrm{diag}(-m,-m,2m)$, shown in Fig.~\ref{fig:kagome} (a), which corresponds to spin space group 65.1.1.1.L~\cite{jiang2024enumeration}. In this configuration, sites 1 and 2 (blue atoms) are related by $C_{2y}$ without spin rotation, whereas the moment at site 3 (red atoms) is not symmetry-related to those at sites 1 and 2. Because opposite spin sublattices are not symmetry-related, it describes a cFiM in which spin degeneracies are generally not protected by spin space group.
     
    Nevertheless, the cFiM model exhibits spin degenerate nodal lines in its band structure. Fig.~\ref{fig:kagome} (b) shows the energy dispersion of the two lowest bands obtained from both the full tight-binding model and the effective Hamiltonian $H_{i,\textrm{eff}}(\bm k)$ in Eq.~(\ref{eqn:Heff}).
    We note that $f_i^\alpha(\bm k)$ can approximately describe the strength of spin splitting between the $i$-th formal Kramers pair at momentum $\bm k$ [see also Supplementary Material (SM.) S1].
    To show the spin degeneracy, we plot the Fermi surfaces of the kagome lattice model for different values of $\mu$ in Fig.~\ref{fig:kagome} (c) in which the magenta lines represent the momenta where the EZF vanishes.
    Notably, for varying $\mu$, the intersections of the two Fermi surfaces consistently lie on these magenta lines. 
    The existence of ZEZF solutions can be understood geometrically in $\mathbb{R}^3$ space, as explained above.
    The intersection of the ZEZF plane and the Hilbert triangle appears as a line segment on the triangle [see Fig.~\ref{fig:kagome} (d)]. By following this line, one can trace the degeneracies across the band.

    Our geometric approach also successfully extends to AMs. For instance, consider $h_m^x=h_m^z=0$ and $h_m^y=\textrm{diag}(m,0,-m)$, for which the system realizes an AM phase: sites 1 and 3 are related by a two-fold spin-space rotation symmetry $C_{2_{110}}$, while site 2 remains invariant.
    As a consequence, along the $C_{2_{110}}$-invariant lines in momentum space, the Bloch eigenstates can be chosen as simultaneous eigenstates of the symmetry operator. This enforces equal weights of the wave function on symmetry-related sublattices, i.e., $|u_{i1}(\bm k)|^2=|u_{i3}(\bm k)|^2$, which directly leads to the vanishing of the EZF, $f_i^y(\bm k)=|u_{i1}|^2-|u_{i3}|^2=0$. From the geometric viewpoint, the images of these eigenstates on the $C_{2_{110}}$-invariant lines lie along the intersection of the Hilbert polygon and the ZEZF plane where the spin degeneracy exists.

    \begin{figure}[ht!]
	\centering
	\includegraphics[width=0.81\linewidth]{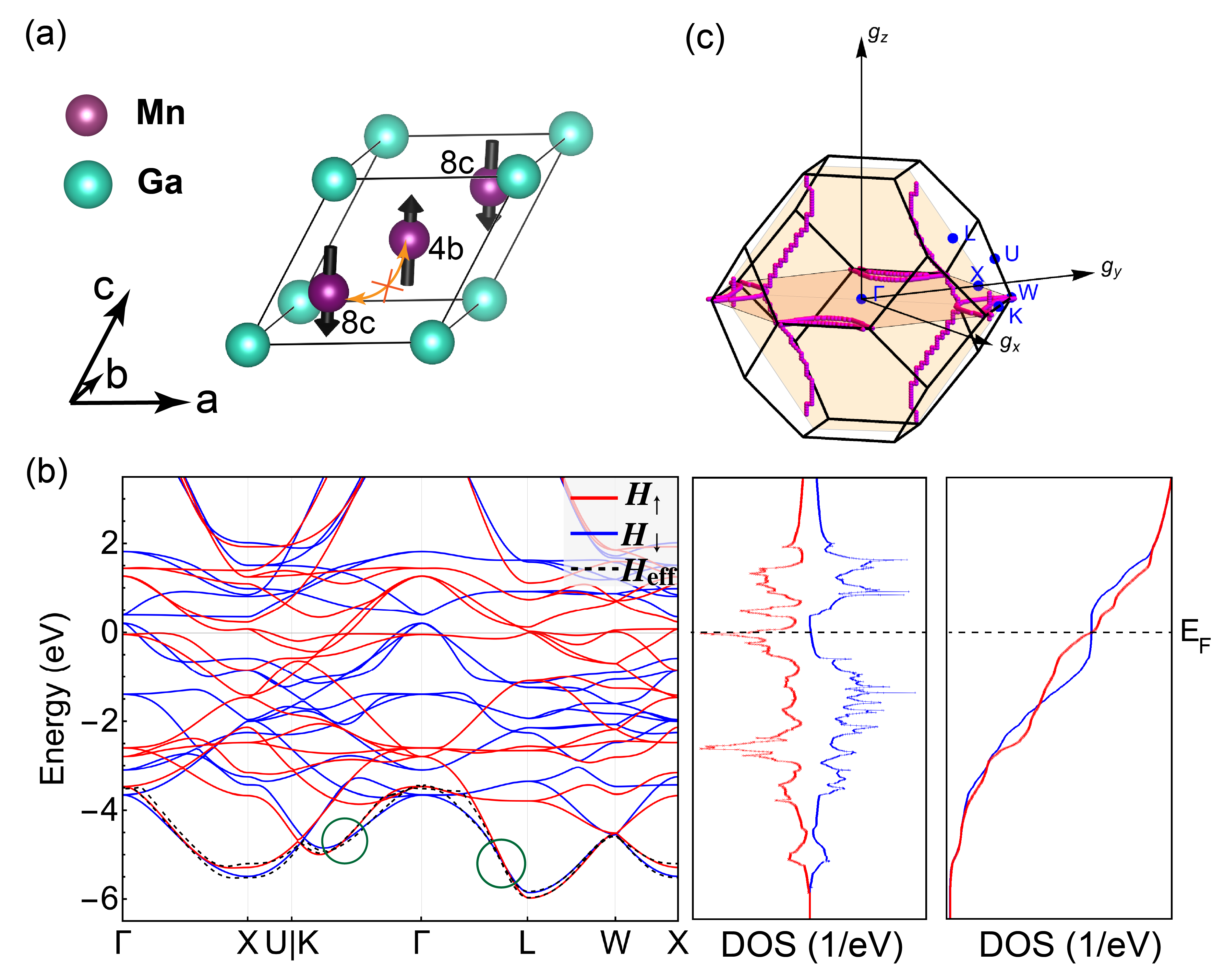}
	\caption{\textbf{Spin degeneracy observed in compensated ferrimagnets Mn$_3$Ga.}
    (a) Crystal structure of Mn$_3$Ga. Arrows indicate magnetic moments.
    (b) Spin-resolved electronic band structure, spin-resolved density of states (DOS), and energy-integrated DOS of Mn$_3$Ga without SOC from DFT calculations. For comparison, the lowest two bands from the effective Hamiltonian are overlaid as dashed black lines.
    (c) ZEZF momenta on the $\Gamma WK$ and $\Gamma LK$ planes (orange planes), indicated by magenta lines, obtained from the $H_\textrm{eff}$ of the two lowest bands. 
    }\label{dft}
    \end{figure}

    \textit{Geometric constraint from compensated magnetization.}\label{constraint_compensation}---Let us describe the geometric meaning of the vanishing net magnetization condition.
    For the kagome lattice model, zero net magnetization implies $\sum_i a^\alpha=\bm a^\alpha \cdot(1,1,1)=0$, where $\bm a^\alpha=(a_1^\alpha,a_2^\alpha,a_3^\alpha)$ and $(1,1,1)$ are the normal vectors of the ZEZF plane and the Hilbert triangle, respectively.
    This orthogonality condition, $\bm a^\alpha\perp(1,1,1)$, implies that the ZEZF plane is perpendicular to the Hilbert triangle. As illustrated in Fig.~\ref{fig:kagome} (d), any plane that is perpendicular to the Hilbert triangle and passes through the origin $(0,0,0)$ must also pass through the center of the triangle, located at $(1/3,1/3,1/3)$.
    In the kagome lattice, the wave function of the lowest-energy parent band at the $\Gamma$ point is $|u_1(\Gamma)\rangle=(1/\sqrt 3,1/\sqrt 3,1/\sqrt 3)$, which maps to the center of the Hilbert triangle due to the three-fold rotation $C_{3z}$ symmetry. Consequently, a spin degeneracy appears at the $\Gamma$ point as shown in Fig.~\ref{fig:kagome} (b). The $K$ and $K'$ points are also invariant under $C_{3z}$. However, because the lowest parent band is degenerate with another parent band, forming a linear Dirac band crossing, the corresponding eigenstates do not map to a single point but instead lie along the incircle of the Hilbert triangle, as illustrated in Fig.~\ref{fig:kagome} (e) [see details in SM. S2].
    More generally, if the nonmagnetic Hamiltonian $H_0(\bm k)$ possesses symmetries that relate all sublattices, these symmetries enforce equal weight of the eigenstate on each sublattice, causing the wave function to map exactly to the center of the Hilbert polygon. In such cases, vanishing net magnetization becomes a \textit{sufficient} condition for spin degeneracy, as the degeneracy is realized geometrically at the intersection between the ZEZF plane and the Hilbert polygon. At high-symmetry momenta where these conditions are satisfied, spin degeneracy can therefore emerge even in the absence of explicit spin-space group symmetries in the full magnetic Hamiltonian $H(\bm k)$.
    This argument generalizes naturally to an $N$-sublattice system. In SM. S3, we provide a couple of examples to illustrate the extended applications of our theory.


    \textit{Spin degeneracy observed in compensated magnets.}\label{motivation}---
    As a representative material example, let us consider the bulk Heusler compound Mn$_3$Ga.
    Mn$_3$Ga/Al with parent space group $Fm\bar{3}m$ are predicted to be half-metallic cFiMs~\cite{wurmehl06}. The magnetic lattice, described by the spin space group 225.1.1.1.L~\cite{jiang2024enumeration}, indicates that the two types of Mn atoms are not symmetry-related. In cubic Heusler compounds, the total magnetic moment $m$ follows the Slater–Pauling rule, an empirical relation that connects the total spin magnetic moment per formula unit with the total number of valence electrons: $m = N_V - 24$, where $N_V$ is the number of valence electrons per unit cell, and 24 corresponds to twice the number of electrons required to fill the minority-spin orbitals to form a closed shell~\cite{Slater,Pauling,KUBLER1984,SP02,bang2019structural}. For the cubic phase of Mn$_3$Ga where $N_V=24$ (21 from three Mn atoms and 3 from Ga), this yields an electron-filling-enforced compensated ferrimagnet with $m = 0$.
    As shown in Fig.~\ref{dft} (a), this band-filling picture manifests in real space: the two Mn atoms at Wyckoff position 8$c$ develop local moments that are antiferromagnetically aligned with the Mn atom at Wyckoff position 4$b$.
    Our density functional theory (DFT) calculations confirm this site-resolved moment cancellation: the central Mn carries –2.880 $\mu_B$, the two neighboring Mn atoms each carry 1.457 $\mu_B$, and the Ga atom has –0.041 $\mu_B$. The total magnetic moment sums to –0.007 $\mu_B$, effectively zero.

    The resulting band structure, displayed in Fig.~\ref{dft} (b), clearly reveals the material’s half-metallic character. Meanwhile, the energy-integrated density of states (DOS), shown in the right panel of Fig.~\ref{dft} (b), reveals that despite the overall imbalance between spin-up and spin-down contributions across the energy range, the two spin channels are exactly matched at the Fermi level. This precise compensation ensures exactly zero total magnetization.
    Furthermore, we observe that below $E = -3$ eV, the exchange interactions between the two spin sectors are relatively weak compared to the bandwidth, as evidenced by the nearly symmetric spin-polarized DOS, and thus our EZF based approach is well suited. This behavior can be attributed to the fact that these states are predominantly derived from the $p$ orbitals of the Ga atoms.
    In this energy range, spin degeneracies are observed between the two lowest bands along the $\Gamma$–$K$, and $\Gamma$–$L$ directions, as highlighted by the green circles in Fig.~\ref{dft} (b), whereas no such degeneracy occurs along $\Gamma$-$X$.
    
    To clarify this behavior, we construct an effective Hamiltonian $H_{\mathrm{eff}}$ by incorporating the EZF into the nonmagnetic Hamiltonian
    [see SM. S4-5 in detail]. While the degeneracy described by $H_{\mathrm{eff}}$ forms a nodal surface, for direct comparison with the DFT band structure we evaluate the ZEZF momenta of the two lowest bands on the $\Gamma WK$ and $\Gamma LK$ planes. The intersections of the nodal surface with these planes give rise to the magenta lines shown in Fig.~\ref{dft} (c), which appear as point degeneracies along the $\Gamma$–$K$ and $\Gamma$–$L$ paths in Fig.~\ref{dft} (b).
    In contrast, no such intersections occur along the $\Gamma$–$X$ direction, and accordingly no degeneracies are found in either the effective model or the DFT band structure.
    

    We note that our EZF framework can also capture symmetry-protected spin degeneracy, as demonstrated in the altermagnetic material RuF$_4$ (see SM. S6).

    \textit{Discussion.}---We have demonstrated that the spin splitting and degeneracy structure of compensated ferrimagnets (cFiMs) can be understood in terms of the geometry of their eigenstates. In the regime where the mean-field exchange coupling to local moments is small compared to the bandwidth, the eigenstates of the nonmagnetic Hamiltonian largely determine the spin splitting in the magnetically ordered phase. In particular, when the net magnetization vanishes, a spin degeneracy emerges within each formal Kramers pair, even in the absence of a symmetry relating the spin-up and spin-down sectors.

    We note that cFiMs have recently attracted considerable attention in spintronics due to their unique combination of antiferromagnetic-like behavior and ferromagnetic-like electrical read/write capabilities~\cite{spintronics}.
    Representative examples include rare-earth–transition-metal alloys~\cite{grundy2006,finley2016spin,caretta2018,cai2020}, where high efficiency of spin-orbit torque has been observed.
    In addition, several Heusler compounds with vanishing net magnetization exhibit half-metallicity~\cite{wurmehl06,kurt2014cubic,half17,he2023fully,midhunlal2018}, making them promising candidates for large tunnel magnetoresistance and high-performance spintronic devices~\cite{jeong2016}.
    Rare-earth iron garnets~\cite{grundy2006,geprags2016,rosenberg2018} provide another important class, featuring magnetization compensation and favorable properties for spin transport. More recently, filling-enforced cFiMs have also been identified~\cite{liuchengcheng}, further expanding the landscape of candidate materials.

    Despite this broad materials platform, a comprehensive theoretical framework for understanding spin degeneracy in cFiMs has remained lacking. Our results provide such a framework by linking spin degeneracy directly to the geometric structure of eigenstates. We anticipate that this perspective will be useful for analyzing spin-dependent phenomena in cFiMs, including the anomalous Hall effect~\cite{Mn3Snroom,2015large,2016giant}, spin–torque responses~\cite{finley2016spin,caretta2018,cai2020}, and the possible emergence of superconductivity driven by complex spin textures~\cite{manna2018}, analogous to proposals in antiferromagnetic systems~\cite{lee2024fermi,Rafael2021}.

    \begin{acknowledgments}
    S.L., Y. Q., B.-J.Y. were supported by Samsung Science and Technology Foundation under project no. SSTF-BA2002-06, National Research Foundation of Korea (NRF) funded by the Korean government (MSIT), grant no. RS-2021-NR060087 and RS-2025-00562579, Global Research Development-Center (GRDC) Cooperative Hub Program through the NRF funded by the MSIT, grant no. RS-2023-00258359, Global-LAMP program of the NRF funded by the Ministry of Education, grant no. RS-2023-00301976.
    \end{acknowledgments}

    \bibliography{ZEZ}

\clearpage

\onecolumngrid

\appendix


\renewcommand{\appendixpagename}{\center\large Supplemental Material for ``\ourtitle''}

\appendixpage

\setcounter{page}{1}
\setcounter{section}{0}
\setcounter{figure}{0}
\setcounter{equation}{0}
\setcounter{table}{0}

\renewcommand{\thefigure}{S\arabic{figure}}
\renewcommand{\theequation}{S\arabic{equation}}
\renewcommand{\thesection}{S\arabic{section}}
\renewcommand{\thetable}{S\arabic{table}}

\noindent
\indent In this Supplemental Material, we provide detailed discussions on the results presented in the main text. The sections are organized as follows.\\\\
\textbf{S1. Method} \hfill \pageref{method}\\\\
\textbf{S2. Comparation of $H(\bm k)$ and $H_\textrm{eff}(\bm k)$ in Kagome lattice} \hfill \pageref{sm:kagome}\\\\
%
\textbf{S3. Role of degeneracies in parent bands}\hfill \pageref{parent_degeneracies}\\\\
\textbf{S4. Extended application of zero net magnetization theory}\hfill \pageref{alter}\\\\
\textbf{S5. ZEZF analysis on ferrimagnet Mn$_3$Ga} \hfill \pageref{mn3Ga}\\\\
\textbf{S6. ZEZF analysis on 2D altermagnet RuF$_4$} \hfill \pageref{RuF4}\\\\


\section{S1. Method}\label{method}
Vienna Ab initio Simulation Package (VASP) package~\cite{KRESSE199615,vasp} based on the density functional theory (DFT) is employed for the first-principles calculations. 
We obtain the crystal structure of Mn$_3$Ga from the paper~\cite{wurmehl06}. 
The generalized gradient approximation (GGA) of the Perdew-Burke-Ernzerhof type is employed for exchange-correlation potential~\cite{pbe96,perdew1998perdew}. 
LSDA~\cite{dudarev1998} is performed to obtain the magnetic electronic band structure of Mn$_3$Ga and RuF$_4$.
The kinetic energy cutoff was set to 500 eV.
For Brillouin zone (BZ) sampling in the self-consistent calculation of Mn$_3$Ga, a $7 \times 7 \times 7$ $k$-point mesh was adopted for self-consistent calculations, and $ 15 \times 15 \times 15$ $k$-point for calculations of the density of states. 
For RuF$_4$, $9 \times 9 \times 1$ $k$-point mesh was adopted for self-consistent calculations. 
The Wannier tight-binding model  for Mn$_3$Ga  is constructed based on $d$ orbitals of Mn and $p$ orbitals of Ga $p$-orbitals, while that for RuF$_4$ is constructed from Ru $d$ orbitals, by the Wannier90 package~\cite{mlwf}.

\section{S2. Comparation of $H(\textbf{k})$ and $H_\textrm{eff}(\textbf{k})$ in Kagome lattice}\label{sm:kagome}

    \begin{figure}[!h]
	\centering
	\includegraphics[width=0.5\linewidth]{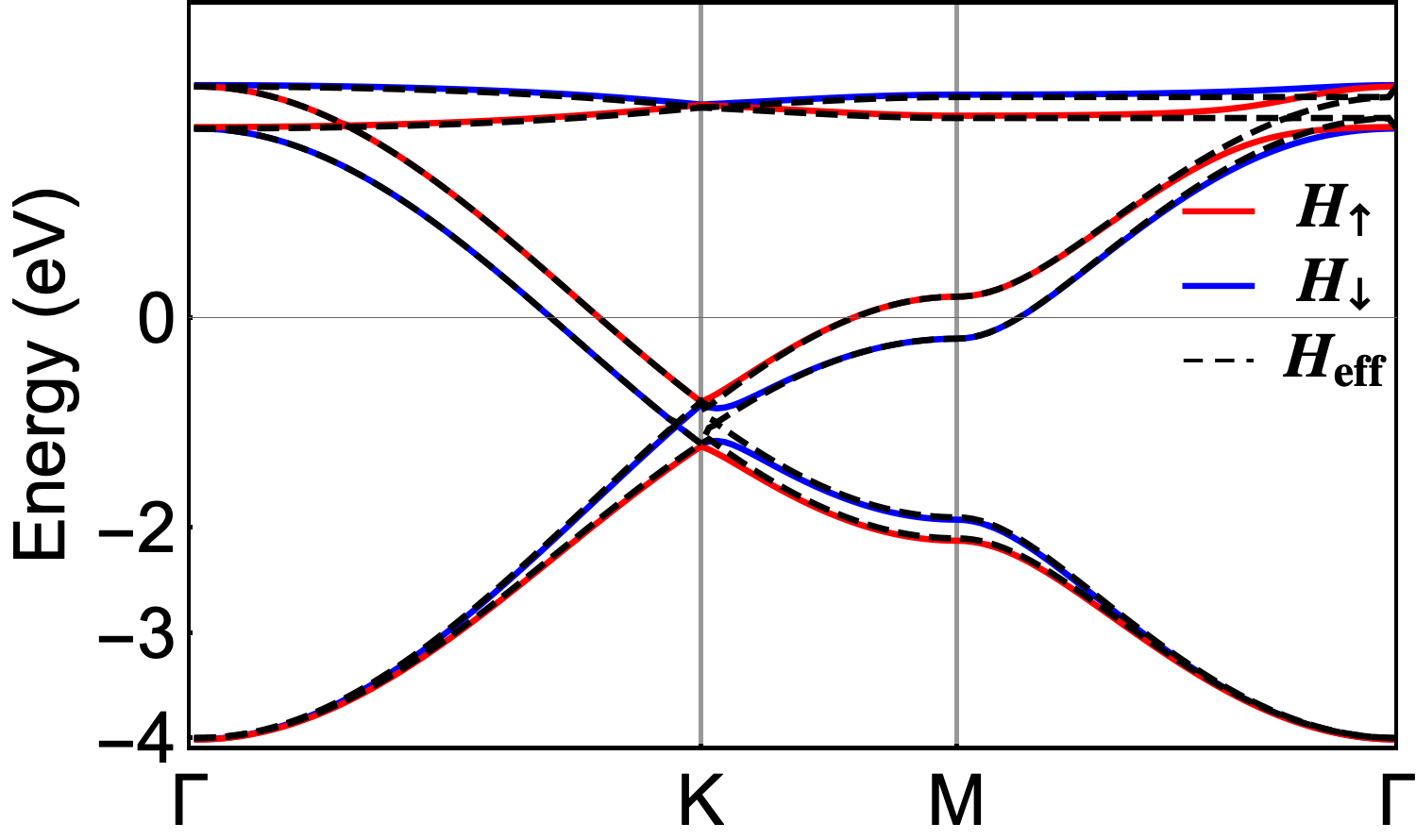}
	\caption{\textbf{Comparation of band structure from $H(\textbf{k})$ and $H_\textrm{eff}(\textbf{k})$.} 
    Red (Blue) band structures are for spin-up (spin-down) bands of $H(\textbf{k})$, and black bands are the band structure of the effective Hamiltonian $H_\textrm{eff}(\textbf{k})$. 
    }\label{degeneracy}
    \end{figure}

Let's recall the form of Hamiltonian $H(\bm k)=H_0(\bm k)\otimes\sigma_0+H_m-\mu I_{2N}$.
$H_0(\bm k)$ is the kinetic part of $H(\bm k)$ and describes the electron hopping.
$H_m=\sum_{\alpha=x,y,z} h_m^\alpha\otimes\sigma_\alpha$ describes the coupling due to local spin moments at sublattice sites.
As we mentioned in the main text, the effective Zeeman field (EZF) is an
approximation that is valid for relatively small spin splitting compared to the separation between parent bands.
As shown in Fig.~\ref{degeneracy}, the band structure from $H_\textrm{eff}(\bm k)$ well capture the global band structure of $H(\bm k)$.
However, there are some discrepancies near the momenta $K$ point along $M-K$ and $\Gamma$ point along $M-\Gamma$ due to the degeneracy of the parent band.
From the result, we found that the $H_\textrm{eff}(\bm k)$ from the effective Zeeman field theory underestimates the band gap between different formal Kramers pairs when magnetic exchange interaction is considered. 
This can be understood because we neglect some coupling terms between different formal Kramers pairs originating from different parent bands in our approximation process.
Nevertheless, within each formal Kramers pair, EZF theory can still capture the most important features of spin degeneracy, as we have discussed in the main text.

\section{S3. Role of degeneracies in parent bands}\label{parent_degeneracies}
    Since the EZF is an approximation valid when the spin splitting is small compared to the interband gap of parent bands, deviations between $H(\bm k)$ and the effective Hamiltonian may arise near the degeneracies of parent bands. For instance, in the kagome lattice, as shown in Fig.~\ref{fig:kagome} (b) and Fig.~\ref{kagomesm}, the mismatch between the dispersions obtained from $H(\bm k)$ and $H_{\textrm{eff}}(\bm k)$ becomes larger near the $\Gamma$ and $K$ points.
    Despite the limitation, the EZF analysis at and near the degeneracies of parent bands still provides useful insights.
    At a degeneracy, the orthonormal basis for the eigenstates of $H_0(\bm k)$ is not uniquly defined. Let us consider an $N_d$-fold degeneracy at $\bm k_d$. Starting from an arbitrary choice of basis for the degenerate eigenstates $|u_i(\bm k_d)\rangle$ ($i=1,2,...,N_d$), infinitely many equivalent bases can be generated by linear combination. When mapped to the space of $V_i$, the set of such eigenstates forms an ($N_d-1$)-dimensional closed manifold in $\mathbb R^N$. As an example, the Dirac nodes at the $K$ ($K'$) point in the band structure of the kagome lattice  an incircle on the Hilbert triangle, as shown in Fig.~\ref{fig:kagome} (e). This can be shown as follows: one possible choice of two orthonormal eigenstates of $H_0(\bm k)$ at $K$ ($K'$) is $(1/\sqrt2,-1/\sqrt2,0)$ and $(-1/\sqrt6,-1/\sqrt6,-2/\sqrt6)$. Their linear combination can be written as $\cos\theta(1/\sqrt2,-1/\sqrt2,0)+\sin\theta(-1/\sqrt6,-1/\sqrt6,-2/\sqrt6)$ where $\theta$ is real. Mapping this to the $V_i$ space yields $(1+\cos2\theta-\sqrt 3\cos\phi\sin2\theta,1+\cos2\theta+\sqrt 3\cos\phi\sin2\theta,2\sin^2\theta)/3$, which corresponds to the points on the blue circle in Fig.~\ref{fig:kagome} (e). The $V_i$ images of eigenstates at momenta near the Dirac nodes lie close to this incircle. The area enclosed by the incircle ($\pi/6$) is larger than the half of the area of the Hilbert triangle ($\sqrt 3/4$). 
    This implies that the ZEZF planes are highly likely intersect with the incircle, making spin degeneracies near the $K$ ($K'$) points.
    We emphasize that for any magnet with zero net magnetization, the corresponding ZEZF plane passes not only through the incircle spanned by the degenerate wavefunctions at $K$ and $K'$, but also through the center of the Hilbert triangle, as illustrated in Fig.~\ref{fig:kagome} (d,e). Consequently, as shown in Fig.~\ref{fig:kagome} (c), the two magenta ZEZF-momenta lines connecting the $K$ and $K'$ points intersect at the $\Gamma$ point.
    These two ZEZF-momenta lines simultaneously connect the $K$ ($K'$) points and the $\Gamma$ point, forming two continuous line segments in the Brillouin zone.
    This connectivity ensures spin degeneracy at $\textit{any}$ chemical potentials, in sharp contrast to magnets without zero net magnetization, where such degeneracy appears only within limited chemical potential ranges.

\section{S4. Extended application of zero net magnetization theory}\label{alter}

    In this section, we now show that our theory provides a general framework for understanding spin degeneracy in all types of magnetic systems [see Fig.~\ref{nearferro} (a-c)]. 
    In this section, we present four additional examples to illustrate extended applications of our theory and examine the consequences when the zero net magnetization condition is violated.
    We also demonstrate that spin degeneracy can persist under certain conditions, even in the presence of symmetry-breaking perturbations.

\begin{figure}[h!]
	\centering
	\includegraphics[width=1.0\linewidth]{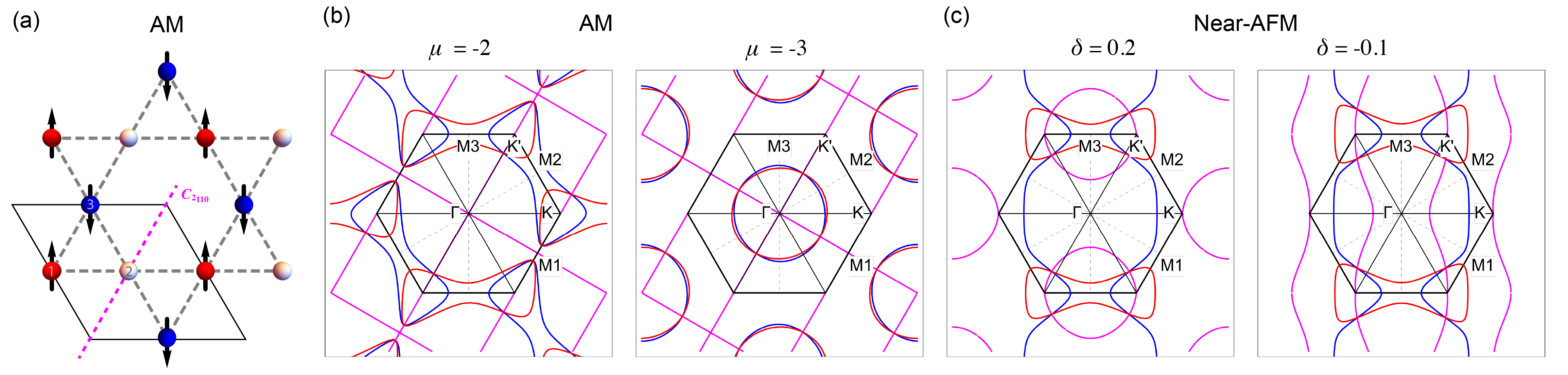}
	\caption{\textbf{Fermi surfaces and ZEZF momenta of AM and near-AFM.}
        (a) AM lattice model with $h_m^x=h_m^z=\textrm{diag}(0,0,0)$ and $h_m^y=\textrm{diag}(m,0,-m)$ at $m=0.2$.
        (b) Fermi surface and ZEZF momenta for AM. Left and right panels for  $\mu=-2$ and $\mu=-3$, respectively.
        (c) Fermi surfaces and ZEZF momenta for near-AFM with $h_m^x=h_m^z=\textrm{diag}(0,0,0)$ and $h_m^y=\textrm{diag}(-m,-m,2m(1+\delta))$ at $m=0.2$ and $\mu=-2$. Left and right panels for $\delta=0.2$ and $\delta=-0.1$, respectively.
    }\label{kagomesm}
    \end{figure}

    {\bf Altermagnet.}--
    Let us first discuss a altermagnet (AM) with $h_m^x=h_m^z=\textrm{diag}(0,0,0)$ and $h_m^y=\textrm{diag}(m,0,-m)$ as shown in Fig.~\ref{kagomesm} (a), which is described by spin space group 65.1.2.5.L~\cite{jiang2024enumeration}. In this configuration, the spin-up and -down sectors are related by the operation $\{M_x||C_{2_{110}}\}$, where $M_x$ and $C_{2_{110}}$ are operations that act on spin and lattice space, respectively. This symmetry classifies the system as an altermagnet. In such cases, we expect spin degeneracy to emerge at momenta satisfying $\mathbf{k}= C_{2_{110}} \mathbf{k}$. Indeed, our calculations of the Fermi surface and zero effective Zeeman field (ZEZF) solutions confirm this expectation, as illustrated in Fig.~\ref{kagomesm} (b). The magenta lines indicate the ZEZF momenta coinciding with high symmetric momenta where $\mathbf{k}= C_{2_{110}} \mathbf{k}$.

    \textbf{Near-antiferromagnets}: As we discussed in the main text, the ZEZF planes always intersect with Hilbert polygon for antiferromagnet (AFM), AM, and compensated ferrimagnet (cFiM) due to zero net magnetization. Here, we introduce the concept of near-antiferromagnets (near-AFM).
    A system is classified as a near-antiferromagnet (or antiferro-like magnet) if, for all spin components $\alpha$, at least one diagonal element $a_n^\alpha$ is negative. In such cases, the ZEZF planes generally intersect with the Hilbert polygon [see Fig.~\ref{nearferro} (c)], indicating a high likelihood of spin degeneracy. 
    The near-AFM can be viewed as a perturbation of the antiferromagnet (AFM), where the strict condition of zero net magnetization is slightly broken.
    As we will show in detail below, spin degeneracy can still persist even if the zero net magnetization condition is slightly violated.

Fig.~\ref{kagomesm} (c) shows two cases with different values of the parameter $\delta$. Compared to the cFiM case shown in the middle panel of Fig.~2 (c), several interesting features emerge.
At first glance, the Fermi surfaces in the cFiM and near-AFM cases appear quite similar. However, in the cFiM case, two ZEZF-momenta lines pass through both the $\Gamma$ and $K$ points and intersect at $\Gamma$. In contrast, for the near-AFM case, these ZEZF-momenta lines are no longer connected and do not pass through the $\Gamma$ point. Instead, they either form two closed loops passing through the $K/K'$ points and encircling the $M$ points, or form two disconnected lines passing through $K/K'$. 

Interestingly, the connectivity difference between near-AFM and cFiM can be easily interpreted by wavefunction geometric way.
As discussed in the main text, the condition of zero net magnetization implies that the normal vector of the EZF plane, denoted as $\mathbf{v}_1$, is perpendicular to the vector (1,1,...,1). Consequently, the ZEZF plane, which contains the origin (0,0,...,0) and has normal vector $\mathbf{v}_1$, always intersects the center of the Hilbert triangle.
In particular, the eigenstate at the $\Gamma$ point of a cFiM in the kagome lattice is mapped to the center of the Hilbert triangle. Therefore, the ZEZF-momenta lines of a cFiM necessarily pass through the $\Gamma$ point.
In contrast, for a general near-AFM system with nonzero net magnetization, the normal vector of the ZEZF plane, denoted as $\mathbf{v}_2$, is no longer perpendicular to (1,1,...,1).
Consequently, since $\mathbf{v}_2$ deviates from $\mathbf{v}_1$, the ZEZF plane defined by normal vector $\mathbf{v}_2$ and containing the origin no longer intersects the $\Gamma$ point. 
On the other hand, the eigenstates at the $K/K'$ points of the kagome lattice are mapped onto the incircle of the Hilbert triangle, as illustrated in Fig.~2 (e). As a result, the ZEZF planes in both cases intersect the incircle, and the corresponding ZEZF-momenta lines necessarily pass through the $K/K'$ points.

\begin{figure}[h!]
	\centering
	\includegraphics[width=0.6\linewidth]{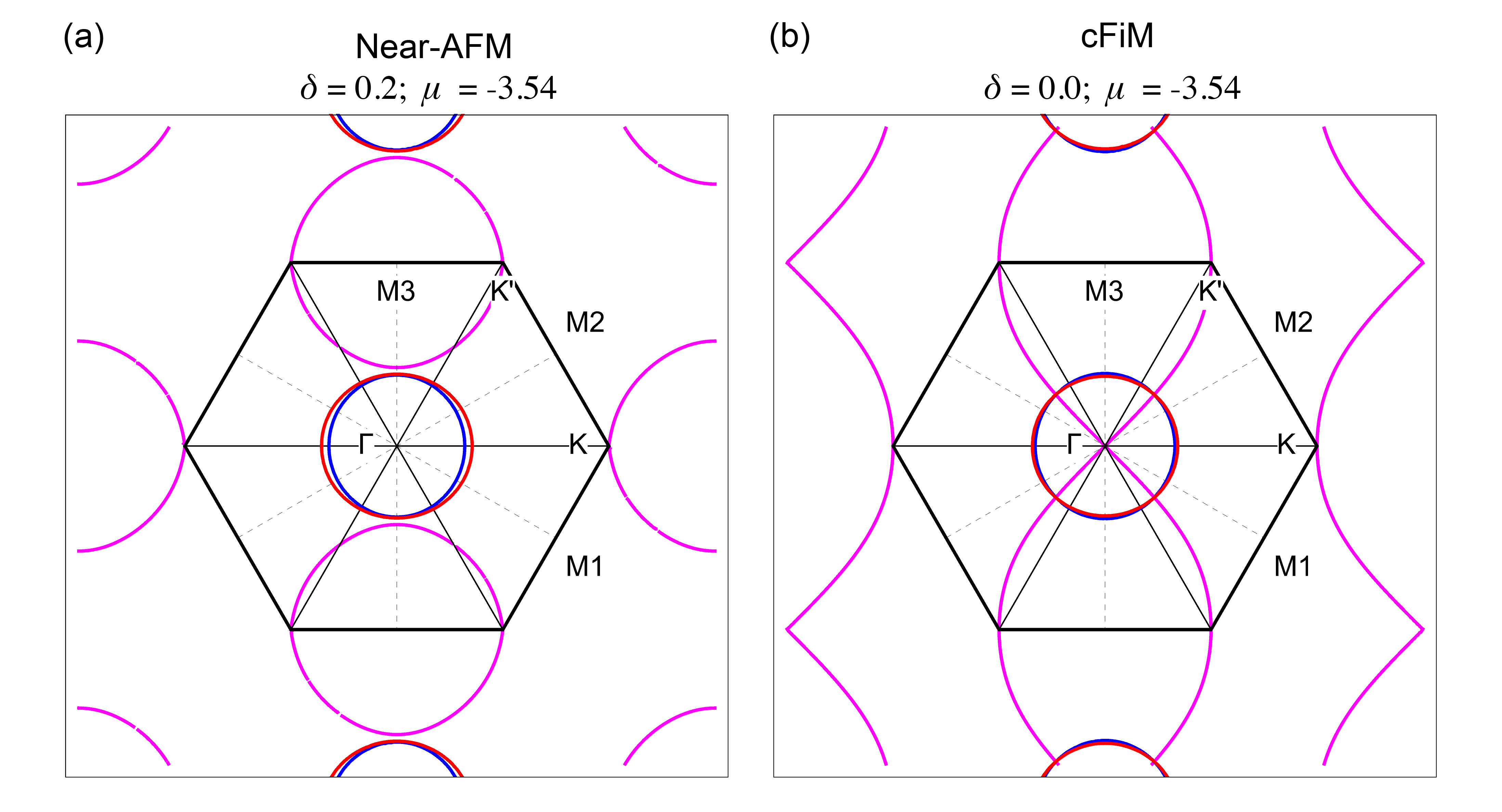}
	\caption{\textbf{Fermi surfaces and ZEZF momenta of near-AFM and cFiM.}
        Fermi surface and ZEZF momenta for $h_m^y=h_m^z=\textrm{diag}(0,0,0)$ and $h_m^x=\textrm{diag}(-m,m,2m(1+\delta))$ at $m=0.2$ and $\mu=-3.54$. (a) and (b) for $\delta=0.2$ (near-AFM) and $\delta=0$ (cFiM), respectively.
    }\label{NearAFM}
    \end{figure}

More importantly, this difference in connectivity reflects a key condition for whether spin degeneracy is preserved. When the ZEZF-momenta lines are connected, as in the cFiM case, spin degeneracy is maintained at any chemical potential $\mu$. When they are disconnected, as in near-AFM, spin degeneracy is generally not guaranteed. This distinction is clearly illustrated in Fig.~\ref{NearAFM}: at $\mu = -3.54$, the spin-up and spin-down Fermi surfaces in near-AFM are not degenerate, whereas they remain degenerate in the cFiM case.

    \textbf{Near-ferromagnets}: If one can find a basis for $h_m^\alpha$s in which every diagonal element $a_n^\alpha$ is positive for any one of $\alpha$s, we call it near-ferromagnet (or ferro-like-magnet). Although the EZF planes of such a system are not parallel to the Hilbert polygon, the ferro-like condition forbids the intersection between the EZF planes and the Hilbert polygon due to the limited size of the Hilbert polygon. So similar to ferromagnets, ferro-like-magnets cannot have degeneracy.

    Interestingly, within our geometric framework for identifying spin degeneracy, a clear phase boundary emerges between ferro-like and antiferro-like magnets. As illustrated in Fig.~\ref{nearferro} (b), the three ZEZF planes in the $N = 3$ case with normal vectors $(1,0,0)$, $(0,1,0)$, and $(0,0,1)$, form this boundary.
    Ferromagnets and ferro-like-magnets that fall entirely on one side of the boundary do not support spin degeneracy.
    In contrast, those intersecting the ZEZF planes, including zero net magnets and antiferro-like magnets, are naturally inclined toward spin-degenerate states.
    This framework provides a new lens for classifying magnetic order: for instance, a canted two-sublattice antiferromagnet with a small net moment may fall into the ferro-like regime.
    Fig.~\ref{nearferro} illustrates this classification, offering a unified geometric view of spin degeneracies across diverse magnetic systems.

\begin{figure}[t!]
	\centering
	\includegraphics[width=0.6\linewidth]{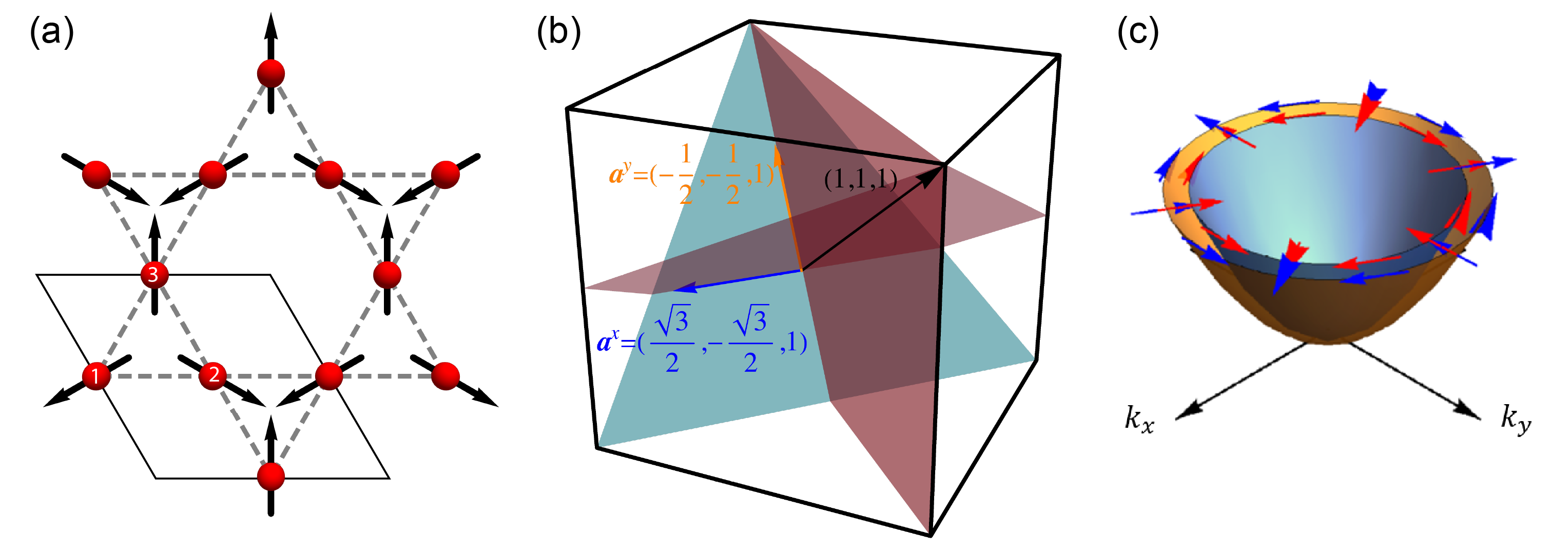}
	\caption{\textbf{Coplanar-non-collinear spin-orbit-free on a kagome lattice.}
        (a) A coplanar-non-collinear magnet on the kagome lattice with  $h_m^x=\textrm{diag}(-\sqrt 3/2,\sqrt 3/2,0)$ and $h_m^y=\textrm{diag}(-1/2,-1/2,1)$. 
        (b) The ZEZF planes for the non-collinear magnet in (a). The normal vectors of the two ZEZF planes are indicated by blue and orange arrows, respectively. The two planes and the Hilbert polygon intersect at a single point rather than along a line.
        (c) The band dispersion and spin texture near the $\Gamma$ point of the lowest band of the non-collinear magnet in (a). The degeneracy between two spin-split bands occurs at $\Gamma$ as a point.
    }\label{collnoncoll}
    \end{figure}

    \textbf{Kagome lattice with non-collinear magnetism:}
    For non-collinear but coplanar configurations, determining spin degeneracy requires solving two independent ZEZF-plane equations. In general non-coplanar magnets, three equations (one per spin component) are required.
    However, particularly when the system has $N=3$ sublattices, the three local spin moments are always coplanar [see Fig.~\ref{collnoncoll} (a)] if their total magnetization is zero, because three vectors whose sum is zero must lie on the same plane. Therefore, in the $N=3$ case with zero net magnetization, only two ZEZF plane equations are sufficient.
    As shown in Fig.~\ref{collnoncoll} (b), the intersection of two ZEZF planes and the Hilbert polygon reduces to a single point. For a wave function mapped to this point, the only remaining degree-of-freedom is the $\mathrm{U}(1)$ phase rotation of each wave-function component, which preserves their magnitudes. However, in the kagome lattice model, infinitesimal changes in momentum inevitably modify the relative amplitudes of the components. 
    As a result, it is impossible to move continuously in momentum space while preserving the ZEZF condition.
    Consequently, the spin degeneracy manifests as isolated nodal points as illustrated in Fig.~\ref{collnoncoll} (c). This is in agreement with the observations in Ref.~\cite{lee2024fermi}.

\section{S5. ZEZF analysis on ferrimagnet Mn$_3$Ga}\label{mn3Ga}
\begin{figure}[!h]
	\centering
	\includegraphics[width=0.85\linewidth]{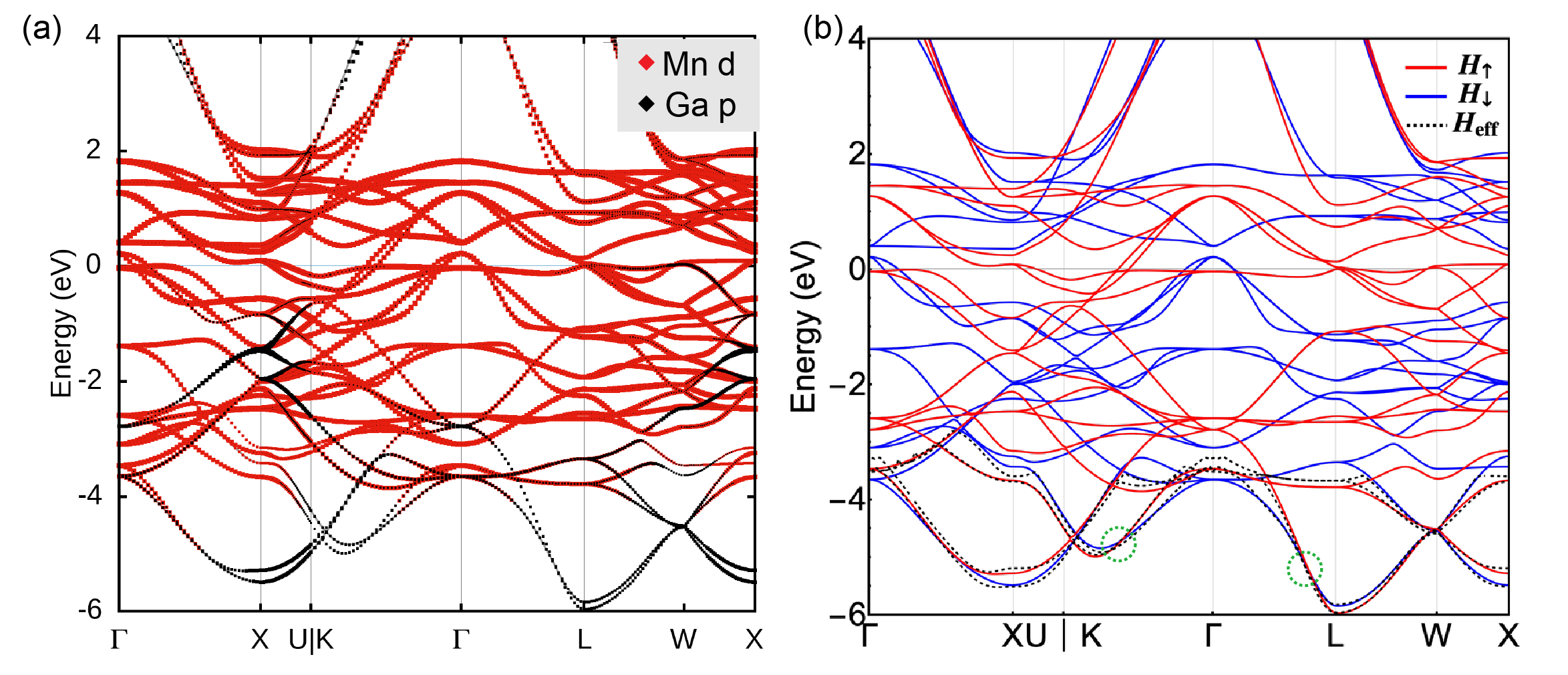}
	\caption{\textbf{Comparison of band structure of Mn$_3$Ga from DFT and effective Hamiltonian $H_\textrm{eff}(\textbf{k})$.} 
    (a) Orbital-resolved band structure of Mn$_3$Ga considering magnetic ordering.
    (b) Red (Blue) band structures are for spin-up (spin-down) bands of $H(\textbf{k})$ directly obtained from DFT, and black bands are the band structure of the effective Hamiltonian $H_\textrm{eff}(\textbf{k})$. 
    }\label{Mn3Ga}
    \end{figure}

In this section, we demonstrate that our ZEZF theory effectively captures the spin degeneracy of the two lowest bands observed along the $\Gamma$–$K$ and $\Gamma$–$L$ directions in Fig.~\ref{dft} (b). We focus on these two bands because their exchange splitting is small compared to their bandwidth, making them well-suited for the ZEZF framework. As illustrated in Fig.~\ref{Mn3Ga} (a), these low-energy bands are primarily derived from the $p$ orbitals of Ga. Since Ga is nonmagnetic, the effective exchange field induced by neighboring Mn atoms leads to only a modest band splitting.
Fig.~\ref{Mn3Ga} (b) demonstrates that the four lowest bands obtained from the effective Hamiltonian successfully capture the spin degeneracies present in the DFT calculations.

\begin{figure}[!h]
	\centering
	\includegraphics[width=0.85\linewidth]{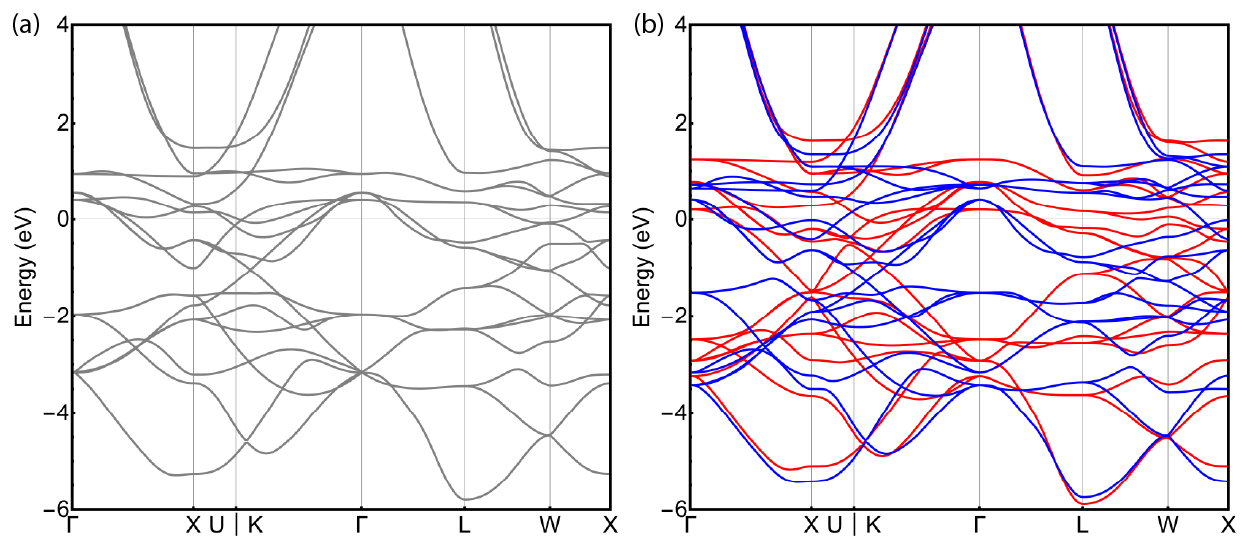}
	\caption{\textbf{Effective band structures of Mn$_3$Ga.}
    (a) Nonmagnetic band structure of Mn$_3$Ga from Wannier and (b) spin-polarized band structure after adding Zeeman term.
    Red (Blue) band structures are for spin-up (spin-down) bands. 
    }\label{Mn3Ga-nomag}
    \end{figure}

Let us now demonstrate how these results are derived from our theoretical framework. We begin by constructing a nonmagnetic Wannier Hamiltonian, $H_0(\mathbf{k})$, for Mn$_3$Ga without considering magnetic ordering. This Hamiltonian includes the Mn $d$ orbitals and Ga $p$ orbitals. The corresponding parent band structure is shown in Fig.~\ref{Mn3Ga-nomag} (a).
To incorporate magnetism, we introduce a Zeeman term and build the full spin-polarized Hamiltonian as 
$
H(\mathbf{k}) = H_0(\mathbf{k}) \otimes \sigma_0 + H_m,
$
where the magnetic term is written as
$
H_m = \sum_{\alpha = x, y, z} h_m^\alpha \otimes \sigma_\alpha,
$
as introduced earlier.
In the case of Mn$_3$Ga, magnetic moments are applied only to the Mn atoms, with Mn atoms at the $4b$ and $8c$ Wyckoff positions carrying opposite spin orientations. The resulting magnetic structure is illustrated in Fig.~\ref{Mn3Ga-nomag} (b). Although the full band structure obtained from this effective Hamiltonian does not exactly match the DFT result, the bands below $E = -3$\,eV show very good agreement in Fig.~\ref{Mn3Ga} (b).
By solving the eigenstates of the parent nonmagnetic Hamiltonian \( H_0(\mathbf{k}) \), we obtain the transformation matrix  
\(
U_0(\mathbf{k}) = \left( |u_1(\mathbf{k})\rangle, |u_2(\mathbf{k})\rangle, \dots, |u_N(\mathbf{k})\rangle \right).
\)  
Using \( U_0(\mathbf{k}) \) and the magnetic field matrix \( h_m \), we compute the EZF for each $i$-th band as  
\(
f_i(\mathbf{k}) = \left[ U_0^\dagger(\mathbf{k})\, h_m\, U_0(\mathbf{k}) \right]_{ii}.
\)  
The resulting four bands, obtained from the effective Hamiltonian \( H_\textrm
{eff}(\mathbf{k}) \) by applying the EZF to the nonmagnetic Hamiltonian \( H_0(\mathbf{k}) \), are shown in Fig.~\ref{Mn3Ga} (b).
Since the band structure obtained from the effective Hamiltonian \( H_{\textrm{eff}}(\mathbf{k}) \) accurately captures the spin degeneracies observed in the DFT calculations, this indicates that these degeneracies arise from vanishing effective Zeeman fields between the formal Kramers bands. 

\section{S6. ZEZF analysis on 2D altermagnet RuF$_4$}\label{RuF4}

\begin{figure}[!h]
	\centering
	\includegraphics[width=0.85\linewidth]{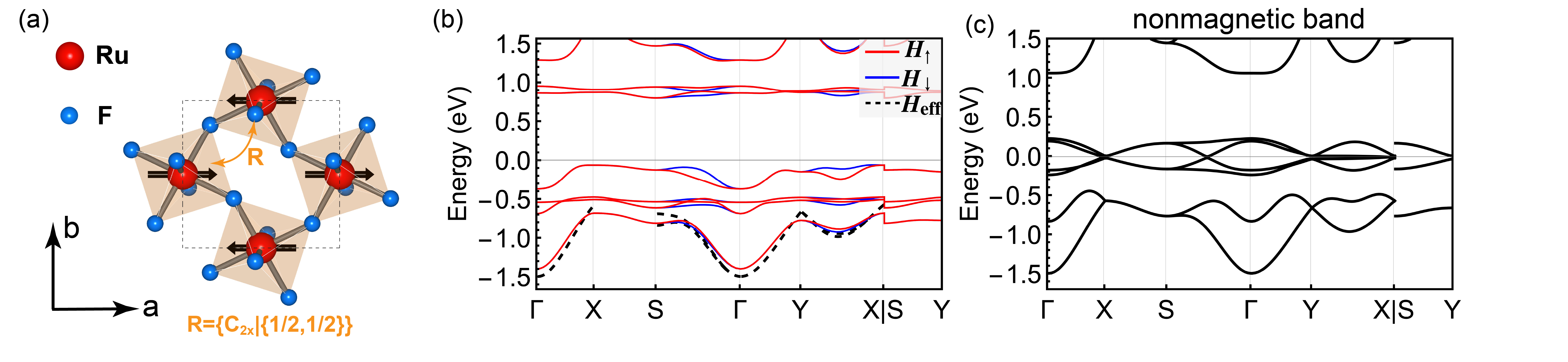}
	\caption{\textbf{Spin degeneracy in band structures of RuF$_4$.}
    (a) Crystal structure of RuF$_4$.
    (b) Spin-resolved band structure of altermagnetic RuF$_4$, together with the effective bands from $H_\textrm{eff}$ constructed from the two lowest bands.
    (c) Band structure of nonmagnetic RuF$_4$.
    }\label{figs:RuF4}
    \end{figure}

Here, we consider monolayer RuF$_4$ as a second example to explicitly demonstrate how our theory accounts for spin degeneracy in another class of compensated magnets--altermagnets. Monolayer RuF$_4$ is reported as a 2D $d$-wave altermagnet~\cite{RuF4_2024,Wang2022}, with spin space group 14.1.2.3.L~\cite{jiang2024enumeration}. As shown in Fig.~\ref{figs:RuF4} (a), two Ru atoms with opposite magnetic moments are related by the spatial operator $R$. Specifically, $R$ can be either $\{C_{2x}|(\tfrac{1}{2},\tfrac{1}{2})\}$, denoting a 180$^\circ$ rotation about the $x$ axis followed by a translation of $(\tfrac{1}{2},\tfrac{1}{2})$, or $\{M_{x}|(\tfrac{1}{2},\tfrac{1}{2})\}$, representing a mirror reflection with respect to the plane perpendicular to $x$ combined with the same translation. These symmetries protect spin degeneracies at $\bm{k}=R \cdot \bm{k}$, ie., along $\Gamma$–$X$, $X$–$S$, $\Gamma$–$Y$, and $Y$–$S$~\cite{RuF4_2024}, consistent with the DFT results in Fig.~\ref{figs:RuF4} (b).

We now examine the validity of our EZF theory. To this end, we start from the nonmagnetic Hamiltonian [Fig.~\ref{figs:RuF4} (c)] and construct an effective Hamiltonian by adding an EZF term. 
The effective Hamiltonian for the lowest band of RuF$_4$ is constructed following the same procedure as for Mn$_3$Ga. 
Specifically, after we first obtain the nonmagnetic band structure from DFT and derive the Wannier Hamiltonian $H_0(\mathbf{k})$ for the nonmagnetic case, we add a Zeeman term into the Hamiltonian. For RuF$_4$, the Zeeman term can be written as $\sigma_x$ for one Ru and $-\sigma_x$ for the other Ru. Thus, the magnetic contribution takes the form $H^x_m= h^x_m \otimes \sigma_x$ with
\[
h^x_m = \mathrm{diag}(1,-1,0,0,0,0,0,0,0,0).
\]
We then project it onto the band basis of $H_0(\mathbf{k})$. By solving the eigenstates of the parent nonmagnetic Hamiltonian $H_0(\mathbf{k})$, we obtain the transformation matrix
$
U_0(\mathbf{k}) = \left( |u_1(\mathbf{k})\rangle, |u_2(\mathbf{k})\rangle, \dots, |u_N(\mathbf{k})\rangle \right).
$
Using $U_0(\mathbf{k})$ and the magnetic field matrix $h_m$, we compute the EZF for each $i$-th band as
$
f^x_i(\mathbf{k}) = \left[ U_0^\dagger(\mathbf{k})\, h_m\, U_0(\mathbf{k}) \right]_{ii}.
$
By adding the EZF to the nonmagnetic Hamiltonian, we obtain the effective Hamiltonian  
$H_\textrm{eff}=H_0(\mathbf{k}) \otimes \sigma_0 + f^x_i \otimes \sigma_x$. 
The band structure obtained from the effective Hamiltonian $H_\textrm{eff}$ for the lowest band is shown as black dashed lines in Fig.~\ref{figs:RuF4} (b).
    
We note that the nonmagnetic band structure is very different from the altermagnetic one due to the large exchange splitting. On the other hand, line degeneracies between pairs of parent bands occur along the $X$–$S$ and $S$–$Y$ directions [see Fig.~\ref{figs:RuF4} (c)]. Despite this limitation, our EZF analysis can still capture some essential degeneracies. As shown by the lowest band in Fig.~\ref{figs:RuF4} (b), the corresponding effective band structure (black dashed lines) obtained by introducing an effective Zeeman field into the nonmagnetic Hamiltonian reproduces the spin degeneracies along $\Gamma$–$X$ and $\Gamma$–$Y$, as well as the spin splitting along $\Gamma$–$S$ and $Y$–$X$.

\end{document}